\documentclass[epjST]{svjour}

\usepackage{graphicx}
\usepackage{epsfig}
\usepackage{dcolumn}
\usepackage{bm}
\usepackage{color}

\def\(({\left(}
\def\)){\right)}
\def\[[{\left[}
\def\]]{\right]}

\newcommand{\beq}{\begin{equation}}
\newcommand{\eeq}{\end{equation}}
\newcommand{\barr}{\begin{eqnarray}}
\newcommand{\earr}{\end{eqnarray}}
\newcommand{\bei}{\begin{itemize}}
\newcommand{\eei}{\end{itemize}}

\begin{document}

\title{Relaxation dynamics of two coherently coupled one-dimensional bosonic gases}

\author{L. Foini \inst{1} \and T.~Giamarchi \inst{2} }

\institute{Laboratoire de Physique Statistique (UMR 8550 CNRS), \'Ecole Normale Sup\'erieure, 
75005 Paris, France \and Department of Quantum Matter Physics, University of Geneva, 24 Quai 
Ernest-Ansermet, CH-1211 Geneva, Switzerland.}

\abstract{
In this work we consider the non-equilibrium dynamics of two tunnel coupled
bosonic gases which are created from the coherent splitting of a one-dimensional
gas. 
The consequences of the tunneling both in the non-stationary regime as
well as at large time are investigated and compared with equilibrium results.
In particular, within a semiclassical approximation, 
we compute correlation functions for the relative phase which are
experimentally measurable and we observe a transient regime displaying oscillations
as a function of the distance. The steady regime is very well approximated by a thermal
state with a temperature independent of the tunneling strength.
}

\maketitle

\section{Introduction}

Thanks to the great achievements in the field of cold atomic experiments
the recent theoretical research interest on the dynamics of closed quantum systems
has increased enormously~\cite{Altman_2105,RSLS_2015,PSSV2011}.
This class of experiments allows one to prepare the system in highly excited states
in a controllable manner, also because of the high tunability of the underlying Hamiltonians,
to monitor their dynamics in isolation from the coupling to any environment and 
to probe the system under appropriate time scales~\cite{BDZ2008,langen2014}.

In this context some relevant questions that can be addressed from the theoretical and the
experimental point of view are connected with the properties of the relaxation
of such an excited state, with the approach of a possible steady regime
and with the emergence or not of a canonical (or a generalized) 
thermal state~\cite{greiner2002,KWW2006,Gring_Science_2012,LEGRSKRMGS2014,schreiber2015}.

The understanding of how thermalization
occurs in isolated quantum systems have actually been investigated long time
ago~\cite{S1994} and is now subject to a revival of interest.
Moreover, the comprehension of the relaxation dynamics of isolated many-body systems is 
of crucial importance also  to have under control the experiments where cold atomic 
systems are designed to probe equilibrium phases as they inevitably evolve 
in absence of coupling to a thermal bath which ensures the thermalization
of the system.

In this dynamical setting a lot of attention has been devoted to the understanding of the role
of integrability and the stationary ensemble that is reached under this 
constraint~\cite{VR_2016,CEF_2011,IDNWBC_2015,CK_2012,CIC_2012,mussardo2013}.
In particular, due to the presence of integrals of motion, many theoretical works have pointed out 
the emergence of many (an extensive number of) temperatures or chemical potentials
that ensure the conservation of local quantities.
Experimentally the existence of many generalized temperatures (or generically Lagrange
multipliers) in a nearly integrable system have been shown in~\cite{LEGRSKRMGS2014}.

However the macroscopic differences between a generalized or a
standard Gibbs ensemble are not always very pronounced and
it is important to understand when this happens and if there are 
constants of motion more constraining than others.

Moreover both from the theoretical and the experimental point of view it 
is natural to ask about the effect of perturbations on such generalized ensembles and 
when a genuine thermalization of the system occurs.
True thermalization is a time dependent phenomenon which might have
very long time scales and is preceded by a transient pre-thermal state.
 In~\cite{berges2004} 
prethermalization was presented as the equilibration at short times of some
thermodynamic quantities while some other being still out-of-equilibrium, 
which is succeeded on a much larger time scale
by the full equilibration of all observables. 
A prethermalized state was experimentally observed in~\cite{Gring_Science_2012}
and interpreted as the decoupling of two types of modes in the system.
From a renormalization group perspective it is 
interesting that integrability breaking interactions that in equilibrium 
might be irrelevant in a non equilibrium unitary dynamics can drive
the system to stationary states with different properties than those of the integrable states 
where eventually a single thermodynamic temperature governs the dynamics~\cite{MG_2011}.

This work is inspired by a set of remarkable experimental and  theoretical achievements 
on the understanding of the relaxation dynamics of quantum many-body systems  
that have been carried on in a long series of works~\cite{Gring_Science_2012,LEGRSKRMGS2014,KISD11,SMLK2013,LGKRS13,LGS_2016,burkov}. 
In these works they considered a one-dimensional gas of bosons which is suddenly split into two one-dimensional gases, sharing
almost the same phase profile.
In all these works the two gases after the splitting were independent
(one controlled way to study the splitting is provided by the ladder system~\cite{FG2015}).
The goal of the present study is to consider the effect of a tunnel coupling term between the two
gases after the splitting is performed.  
We study how this term changes the relaxation of correlation functions that can be directly measured in 
interference measurements and its consequences on the effective temperature
that emerged from the prethermalized state in~\cite{Gring_Science_2012}.

The manuscript is organized as follows: in Section~\ref{Sec_model} and~\ref{Sec_initial_state} 
we present the model and the characterization of the initial state, in section~\ref{Sec_eigenmodes} 
we describe the thermodynamic and dynamic properties
of the eigenmodes, in section~\ref{Sec_dynamics_phase} we discuss the dynamics
of the relative phase of the two gases and we compare with the equilibrium values,
in section~\ref{Sec_fdt} we study two-time quantities and we look at fluctuation-dissipation relations.
Finally in section~\ref{Sec_discussions} we discuss our results and we conclude.

\section{Theoretical setting}\label{Sec_model}

\begin{figure}
  \begin{center}
         \includegraphics[scale=0.3]{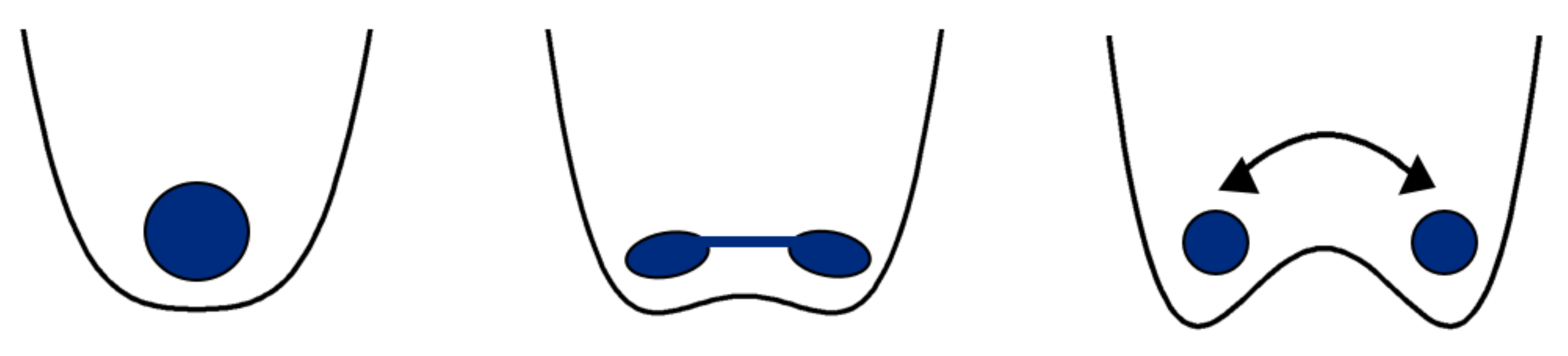}
        \caption{ Schematic figure of the splitting protocol that leads from one single
        gas to two tunnel-coupled tubes.
        } 
       \label{FIG_sketch}
  \end{center}
\end{figure}

The system is prepared as a one-dimensional gas and, at time $t=0$, a 
barrier is grown in order to create two one-dimensional systems sharing
almost the same phase profile~\cite{KISD11,Gring_Science_2012,LGKRS13}.
We suppose that the barrier is grown instantaneously and kept finite. 
In Fig.~\ref{FIG_sketch} we show a sketch of this situation. 

The Hamiltonian of the 
two systems after the splitting is well described within the Luttinger liquid theory by~\cite{giamarchi2004}:
\beq
\displaystyle H = H_{LL}^1 + H_{LL}^2 - \, \frac{t_{\perp}}{2\pi}  \, \int {\rm d} x \, 2  \cos( \theta_2(x) - \theta_1(x)) \ .
\eeq
Here $H_{LL}^\alpha$ is the Luttinger liquid Hamiltonian describing the system $\alpha=1,2$:
\beq
\displaystyle H_{LL}^\alpha =\frac{\hbar u}{2} \int {\rm d} x \, \Big[ \, \frac{\mathcal{K}}{\pi} \, [\nabla \theta_{\alpha}(x) ]^2 \, + \,\frac{\pi}{\mathcal{K}} [n_{\alpha}(x) ]^2 \, \Big] \ ,
\eeq
$u$ is the sound velocity and $\mathcal{K} $ is the Luttinger parameter which encodes the interactions of the system:
$\mathcal{K}=1$ corresponds to hard core bosons while $\mathcal{K}=\infty$ to free bosons.
In the weakly interacting regime realized in the experiment, in terms of the microscopic parameters of the gas
one has $\mathcal{K} = \hbar\pi \sqrt{\frac{\rho}{M g}}$ and $u= \sqrt{\frac{g \rho}{M}} $, where $\rho$ is the
 density of atoms in each of the tubes, $M$ their mass 
and $g$ the strength of their effective one dimensional interaction.

The operators $\theta(x)$ and 
$n(x)$ representing respectively the 
phase of the bosonic field and the fluctuation of its density are canonically conjugated:
$[n(x),\theta(x')]=i \delta(x-x')$.
The cosine term originates from the tunneling operator 
$\psi_1^{\dag}(x)\psi_2(x) + {\rm h.c.} = 2 \, \rho   \cos( \theta_2(x) - \theta_1(x))$,
where we used that $\psi_\alpha(x) \simeq \sqrt{\rho} \, e^{i \theta_\alpha(x)}$ and
 $\rho$ is the density. Therefore one can set $t_{\perp} = \,\hbar \,J \rho$, where $\hbar J$
 is the energy scale defined by the potential barrier. 
 
 The dynamics of this system can be studied by introducing symmetric and antisymmetric variables 
 $\theta_{A/S} = \theta_1 \mp \theta_2$ and $n_{A/S} = (n_1 \mp n_2)/2$. 
 Indeed under the assumption that the two systems are identical, symmetric and antisymmetric modes 
  decouple and in terms of those variables one has $H = H_{LL}^S + H_{SG}^A$
with $H_{SG}^A $ the Sine-Gordon Hamiltonian for the antisymmetric modes:
\beq\label{HA}
\displaystyle H_{SG}^A =  \frac{\hbar u}{2} \, \int {\rm d} x \, \Big[ \, \frac{K}{\pi} [\nabla \theta_A(x)]^2 + \frac{\pi}{K} [n_A(x)]^2 \, \Big] - \, \frac{t_{\perp}}{2\pi}  \, \int {\rm d} x \, 2 \, \cos \, \theta_A(x) \ .
\eeq
Here $K= \mathcal{K}/2 $. 
The series of experiments~\cite{KISD11,Gring_Science_2012,LGKRS13} are compatible with $K\sim 30$ which implies that they are in
the weakly interacting regime.
In the following we will focus only on the antisymmetric part of $H$. 
In order to diagonalize the Hamiltonian and to study its dynamics we will make the following semiclassical
approximation:
\beq\label{HAsc}
\displaystyle H_{SC}^A =  \frac{\hbar u}{2} \, \int {\rm d} x \, \Big[ \, \frac{K}{\pi} [\nabla \theta_A(x)]^2 + \frac{\pi}{K} [n_A(x)]^2 \, \Big] + \, \frac{t_{\perp}}{2\pi}  \, \int {\rm d} x \,   [ \theta_A(x) ]^2 \ ,
\eeq
which corresponds to develop the cosine up to second order around its minimum.
This approximation should be well justified in the
weakly interacting regime of the experiment.
The Hamiltonian~(\ref{HAsc}) can be studied expanding the operators $\theta_A$ and $n_A$
in the base of bosonic operators $\{b_p,b_p^{\dag}\}$ that diagonalize the Luttinger liquid
Hamiltonian, corresponding to the case $t_{\perp}=0$:
\beq\label{Initial_correlations}
\begin{array}{ll}
\displaystyle n_A(x) = \frac{1}{\sqrt{L}} \sum_{p \neq 0} e^{-i p x} e^{- (p \xi_h)^2 /2} \sqrt{\frac{K |p|}{2 \pi}} (b_p^{\dag} + b_{-p}) + \frac{1}{\sqrt{L}} n_0
\\ \vspace{-0.2cm} \\
\displaystyle \theta_A(x) = \frac{i}{\sqrt{L}} \sum_{p \neq 0} e^{-i p x} e^{- (p \xi_h)^2 /2} \sqrt{\frac{\pi}{2 K |p|}} (b_p^{\dag} - b_{-p}) + \frac{1}{\sqrt{L}} \theta_0  \ ,
\end{array}
\eeq
where $n_0$ and $\theta_0$ represent the $p=0$ Fourier transform of the fields, and we
assumed that there is a high energy cut-off of the order of the inverse of the healing length 
$\xi_h = \hbar  u / g \rho$.

Next we perform a canonical transformation from the operators $\{b_p,b_p^{\dag}\}$ to the $\{\gamma_p,\gamma_p^{\dag}\}$:
\beq
\begin{array}{ll}
\displaystyle \gamma_p^{\dag} = \cosh \varphi_p b_p^{\dag} - \sinh \varphi_p b_{-p}
\\ \vspace{-0.2cm} \\
\displaystyle  \gamma_{-p} = \cosh \varphi_p b_{-p} - \sinh \varphi_p b_{p}^{\dag} \ ,
\end{array}
\eeq
with $\tanh 2 \varphi_p = \frac{t_{\perp}}{t_{\perp}+2 \hbar K u |p|^2}$, which 
diagonalizes the Hamiltonian (\ref{HAsc}):
\beq
H_{SC}^A = \sum_{p} \, \omega_p \gamma_p^{\dag} \gamma_p \ ,
\eeq
with $\omega_p = \sqrt{(\hbar u p)^2 + t_{\perp} u \hbar / K} = \sqrt{(\hbar u p)^2+m^2}$ where we set $m^2=t_{\perp} u \hbar / K$.

\section{Initial state}\label{Sec_initial_state}

Following~\cite{Gring_Science_2012,KISD11}, we assume that just after the splitting the system is 
well described by a minimum uncertainty state characterized by the following correlation functions: 
\beq\label{Initial_correlations}
\begin{array}{ll}
\displaystyle \langle n_A(p) n_A(p')  \rangle(t=0) =  \frac{\rho}{2}  \, \delta_{-p,p'} \ ,
\\ \vspace{-0.2cm} \\
\displaystyle \langle \theta_A(p) \theta_A(p') \rangle(t=0) =  \frac{1}{2 \rho} \, \delta_{-p,p'} \ ,
\end{array}
\eeq
resulting in local correlations in real space $\langle n_A(x) n_A(x')  \rangle(t=0) =  \frac{\rho}{2}  \, \delta(x-x')$.
This form of correlation functions should intend that the delta function is 
smeared over the healing length scale $\xi_h$.
The strength of density fluctuations after the splitting (and consequently of phase fluctuations) is chosen
to be proportional to the density itself assuming that in the splitting process particles can go either left of right
with equal probability and the number of particles is large.

\section{Eigenmodes' properties}\label{Sec_eigenmodes}

\subsection{Eigenmodes' energy}

A quantity of particular interest in order to understand the effective ``thermodynamics" of the system
after the quench is the average energy of the system.
In the quadratic approximation one can look at the energy of each mode and this is given by:
\beq\label{Eq_Ek}
\langle E_p \rangle = \frac{\omega_p}{2} \, \Big[\frac{K \omega_p}{\hbar u\pi} \,  \, \langle |\theta_p|^2\rangle(0) + \frac{\hbar u \pi}{K \omega_p}   \langle |n_p|^2\rangle(0) \Big] \ .
\eeq
 If one neglects the initial fluctuation of the phase, setting $\langle |\theta_p|^2\rangle(0) \sim 0$, and 
 assumes that classical equipartition holds, Eq.~(\ref{Eq_Ek}) provides a $p$-independent effective temperature:
 \beq\label{Teff}
 E_k \sim T_{\rm eff} =  \frac{\hbar u \pi \rho}{4 K}  \ .
 \eeq
This reasoning leads to an effective temperature which is also independent of the
dispersion $\omega_p$ and therefore of the tunneling $t_{\perp}$. 
In the following we will show that this effective temperature is a meaningful quantity that
allows to characterize very well the correlation functions of the relative phase.
Let us note that neglecting the contribution from $\langle |\theta_p|^2\rangle(0)$ 
can have an impact on low energy modes if $\frac{K \omega_0}{\hbar u\pi \rho}  =  \frac{\xi_{h} m}{2 \hbar u} \gg 1$.
For high energy modes $p>\xi_{h}^{-1}$ this approximation starts not to be true, however
their contribution is suppressed in the correlation functions.

\subsection{Eigenmodes' dynamics}

Under the semiclassical approximation the dynamics of each mode is decoupled from the other and one can compute the
time evolution of density and phase fluctuations which are given as follows:
\beq\label{Dyn_theta}
\begin{array}{ll}
 \displaystyle 
\displaystyle \langle |\theta_p|^2 \rangle (t)
= \sin^2 \omega_p t \, \Big( \frac{\hbar u\pi}{K  \omega_p} \Big)^2 \, \langle |n_p|^2 \rangle (0) + \,  \cos^2 \omega_p t \, \langle |\theta_p|^2 \rangle (0)  \ ,
\end{array}
\eeq
\beq
\begin{array}{ll}
 \displaystyle 
\displaystyle \langle |n_p|^2 \rangle (t)
&  \displaystyle  = \sin^2 \omega_p t \, \Big( \frac{K  \omega_p} {\hbar u\pi}\Big)^2 \, \langle |\theta_p|^2 \rangle (0) + \,  \cos^2 \omega_p t \, \langle |n_p|^2 \rangle (0)  \ .
\end{array}
\eeq
From these equations one sees that in the massless case $t_{\perp}=0$ the dynamics of the zero mode is different
from that of the other modes and should be treated separately. 
As soon as $t_{\perp}\neq 0$, though, this mode oscillates as the others.

\section{Dynamics of the relative phase}\label{Sec_dynamics_phase}

In the following we will be interested in the expectation values of the  correlation
functions of the relative phase
as it is the most direct observable in interference experiments.
In fact, from the interference pattern of the two gases after time of flight it is possible to extract 
their relative phase at any point $x$ and repeating the measurement many times for all available times 
$t$ it is possible to reconstruct the correlation function~\cite{LGKRS13}:
\beq\label{CthetaA}
 \displaystyle C(x,t) = \langle e^{i (\theta_A(x,t) - \theta_A(0,t))} \rangle =  e^{- \frac12 \langle [ (\theta_A(x,t) - \theta_A(0,t))^2 ]\rangle} \ ,
\eeq
where the second equality holds in the semiclassical limit~(\ref{HAsc}) (see Appendix~\ref{App2}).

\begin{figure}
  \begin{center}
        \includegraphics[scale=0.7]{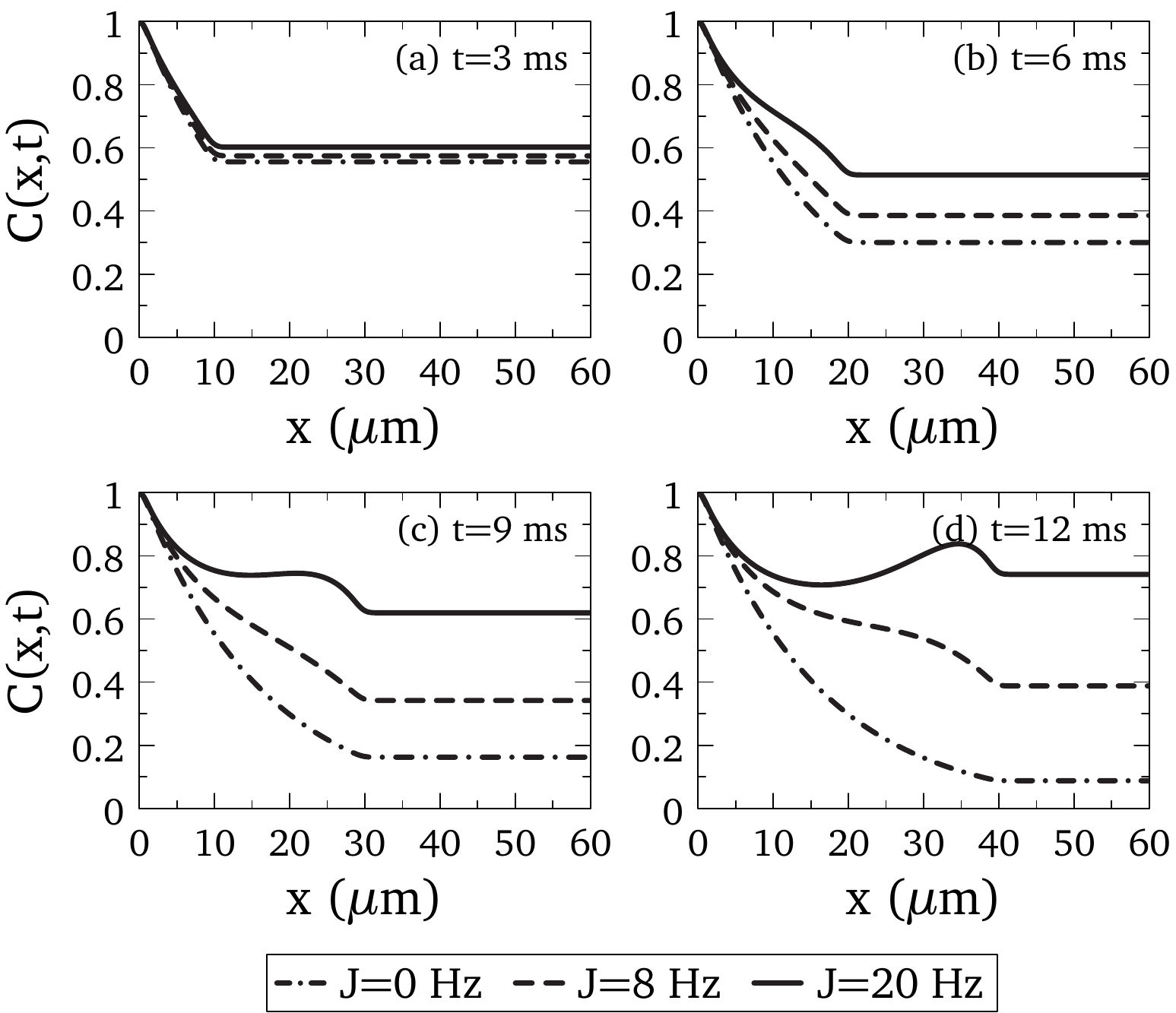}
        \caption{Correlation functions of the relative phase $C(x,t)$ (\ref{CthetaA}), as obtained from~(\ref{CorrCxt}),
        as a function of the distance for different times and coupling $\hbar J = t_{\perp} /\rho$. The four plots correspond
to times $t=3,6,9,12$ ms and in each plot the values $J=0, 8, 20$ Hz are shown with 
dotted-dashed, dashed and solid lines respectively.
While the relaxation of the correlations is independent of the tunneling at short times 
(see Fig.~\ref{FIG1}.a), at larger times the effect of tunneling is
evident in the long range order (see Fig.~\ref{FIG1}.b, c and d).
        }\label{FIG1}
  \end{center}
\end{figure}

Unsing Eq.~(\ref{Dyn_theta}) and (\ref{CthetaA})
we arrive at the following form for the correlation functions:
\beq\label{CorrCxt}
\displaystyle C(x,t) 
= \exp\Big[- \frac{1}{4 K \lambda}  I_1(\tilde{x},\tilde{t},\tilde{m}) -  \frac{\lambda}{4 K} I_2(\tilde{x},\tilde{t},\tilde{m})  \Big] \ ,
\eeq
with 
$\displaystyle \lambda = \frac{K}{\pi \rho \xi_{h}} = \frac12$  and $\tilde{x}=x/\xi_{h} $, $\tilde{t}=t u /\xi_{h}$, $\tilde{m}=\xi_{h} m /\hbar u$:
\beq\label{Def_I1}
\begin{array}{ll}
\displaystyle I_1(\tilde{x},\tilde{t},\tilde{m}) & = \displaystyle  \int_0^{\infty} {\rm d} \tilde{p}   \,e^{- \tilde{p}^2} \, \frac{1}{\tilde{m}^2+\tilde{p}^2} \, \Big(1 - \cos(\tilde{p}\tilde{x}) \Big) 
\Big[ 1 - \cos( 2 \tilde{t} \sqrt{\tilde{m}+\tilde{p}^2}) \Big]  \ ,
\end{array}
\eeq
\beq\label{Def_I2}
\begin{array}{ll}
\displaystyle I_2(\tilde{x},\tilde{t},\tilde{m}) & = \displaystyle  \int_0^{\infty} {\rm d} \tilde{p}  \,e^{- \tilde{p}^2}  \, \Big(1 - \cos(\tilde{p}\tilde{x}) \Big) \,
\Big[ \cos( 2 \tilde{t} \sqrt{\tilde{m}^2+\tilde{p}^2}) + 1 \Big] \ , 
\end{array}
\eeq
where 
$e^{-p^2}$ is a function regularizing the integral.

In Fig.~\ref{FIG1} we show the correlation functions, as obtained from~(\ref{CorrCxt})
for different times and coupling $\hbar J = t_{\perp} /\rho$. In particular the four plots correspond
to times $t=3,6,9,12 \ ms$ and in each plot the values $J=0, 8, 20 \ Hz$ are shown with 
dotted-dashed, dashed and solid lines respectively.
Fig.~\ref{FIG1} and the following show the results obtained for the correlation function 
$C(x,t)$ in (\ref{CorrCxt}) computed with
       $\rho=40\,\mu m^{-1}$ , $\xi_h \simeq 0.45 \ \mu m$,
       $K = \frac{\pi}{2} \xi_h \rho  \simeq 30$ and 
       $u   \simeq 1.6 \, \mu m / ms$.
While the relaxation of the correlations is independent of the tunneling at short times 
up to a certain extent (see Fig.~\ref{FIG1}.a), at larger times the effect of tunneling is
evident in the long range order (see Fig.~\ref{FIG1}.b, c and d).

\begin{figure}
  \begin{center}
        \includegraphics[scale=0.45]{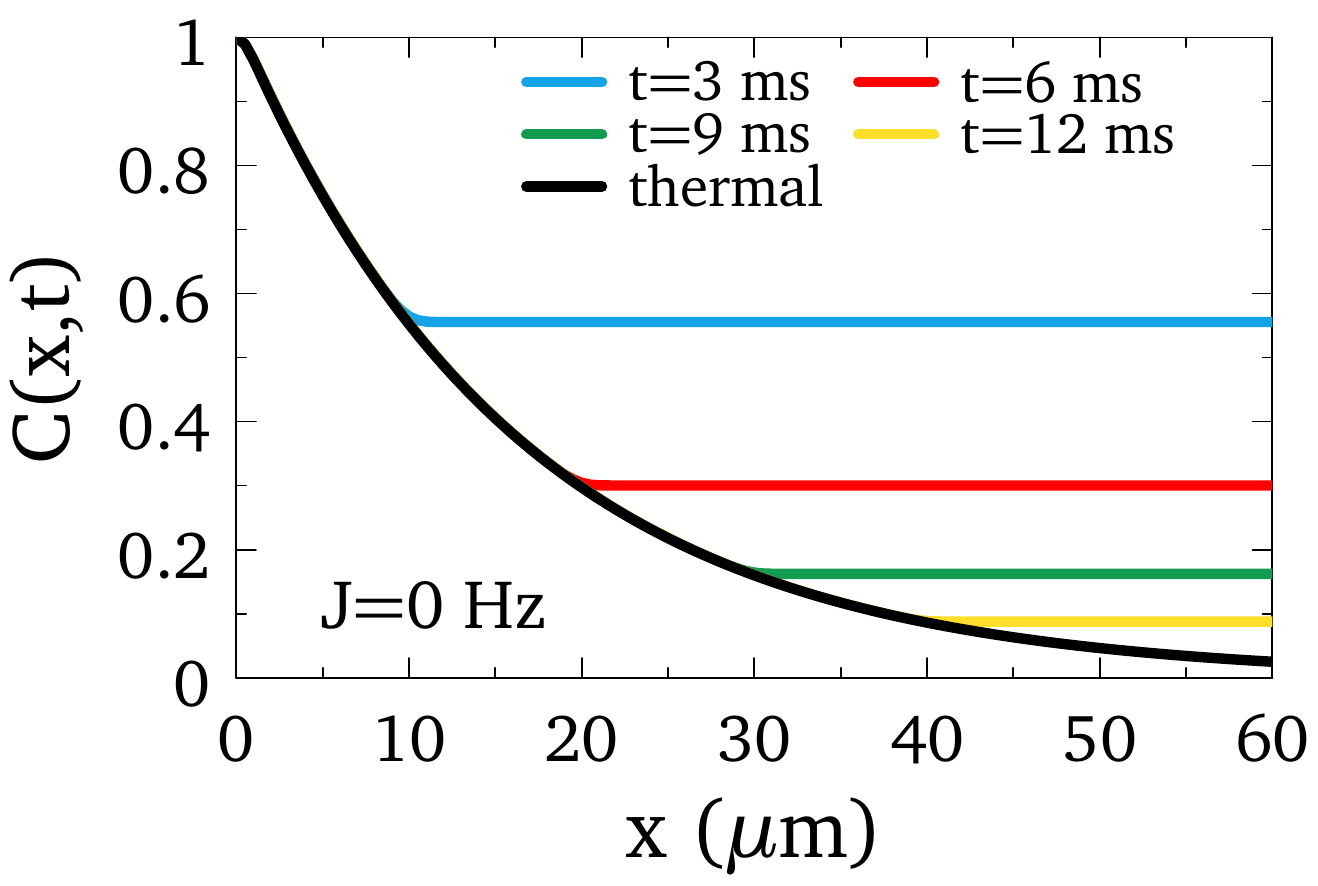}
   \includegraphics[scale=0.45]{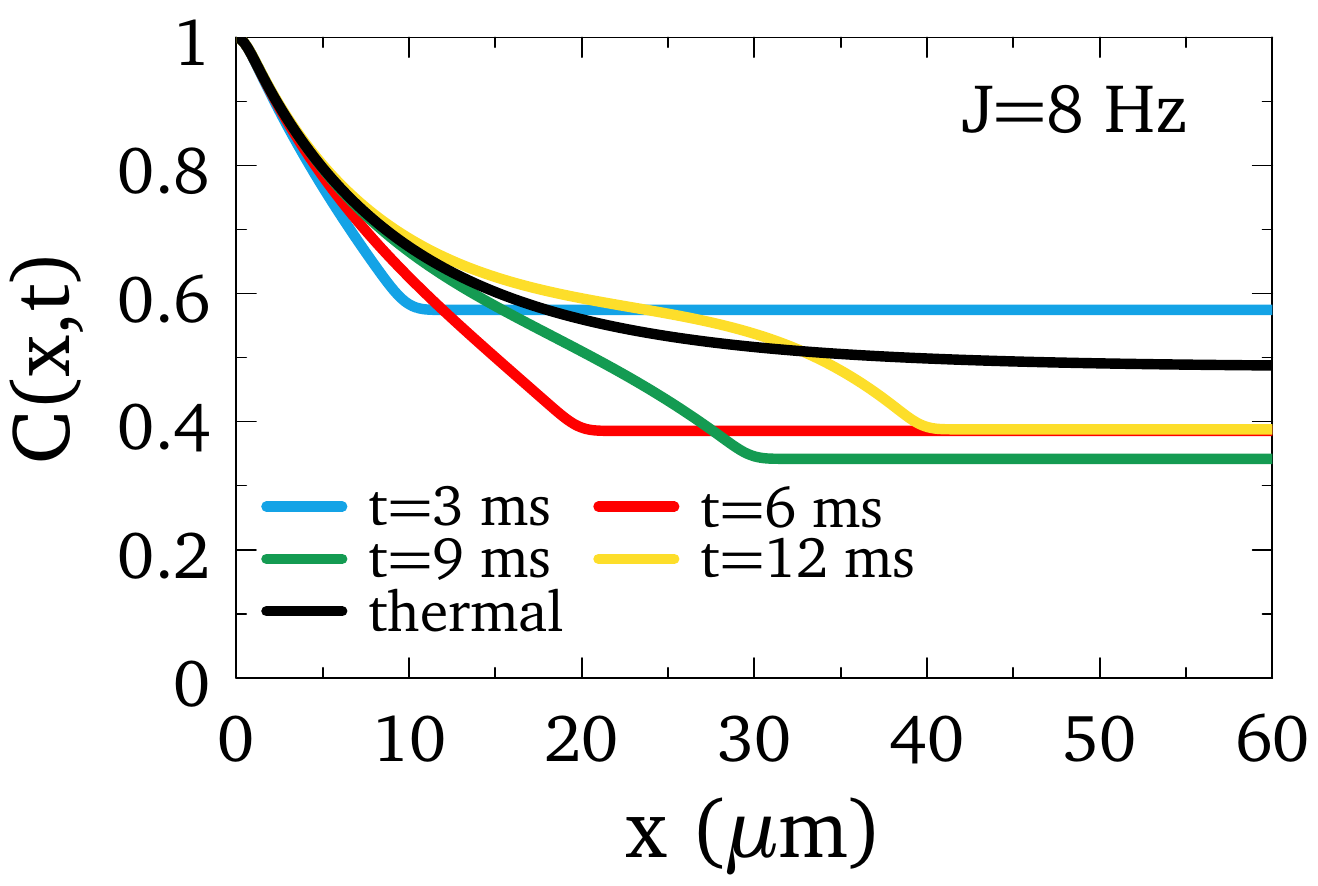}
      \includegraphics[scale=0.45]{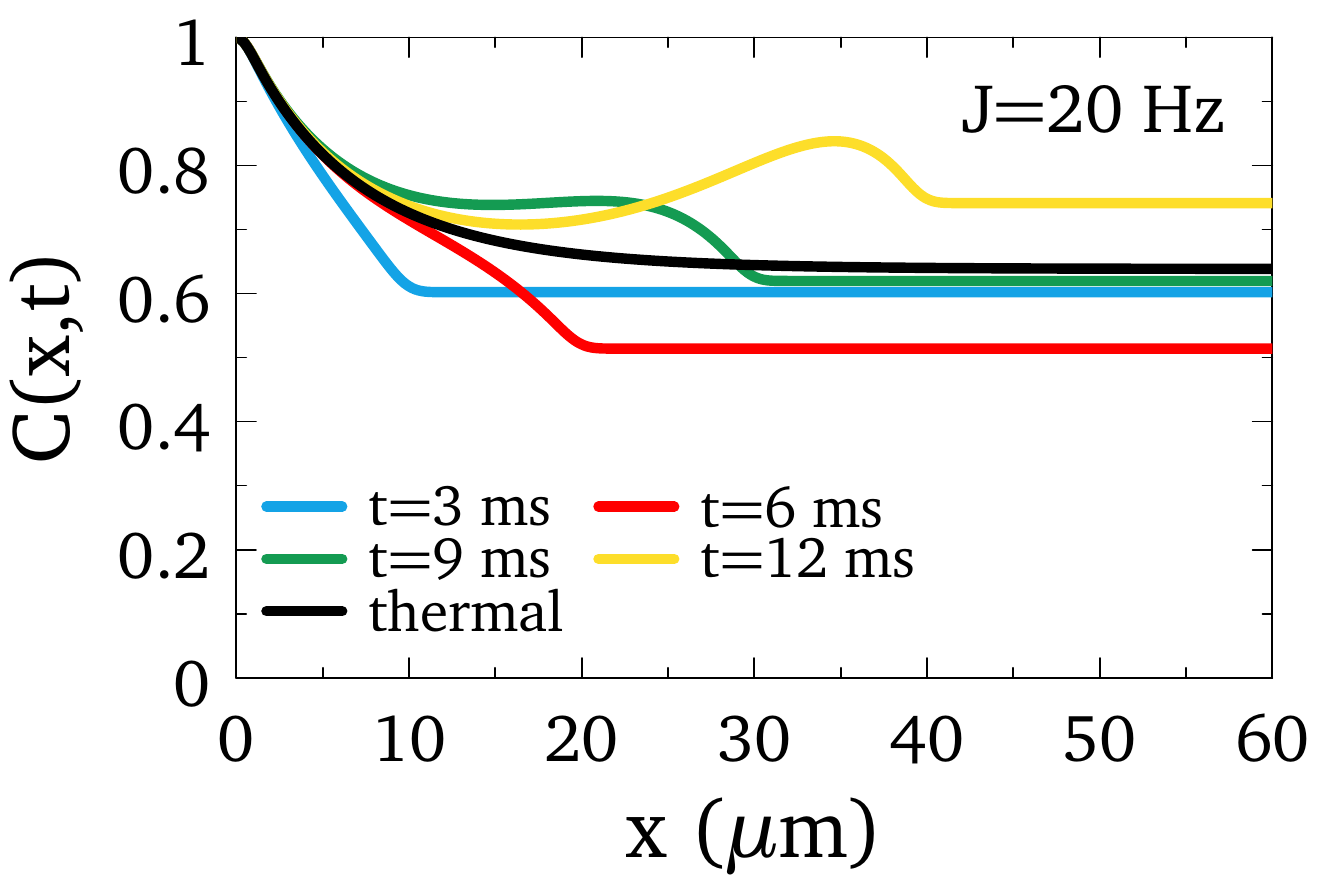}
        \caption{Spatial correlation functions for different times (coloured lines) compared
        with the thermal ones at inverse temperature $\beta_{\rm eff} =  \frac{4 K}{\rho \pi \hbar u}$ (black lines). Blue, red, green and yellow lines correspond respectively
        to $t=3,6,9,12$ ms.
        The three plots refer to
        different values of $J=0, 8,20$ Hz. 
        } 
       \label{FIG2}
  \end{center}
\end{figure}

In Fig.~\ref{FIG2} we show spatial correlation functions grouped according to the strength
of the tunneling $J = 0, 8, 20 \ Hz$ and for different times. 
As in the case without tunneling one sees a light cone in the correlation functions propagating
linearly at $x_c=2 u t$. The decay towards the stationary regime is different,
characterized in particular by an oscillating behavior (which is more visible at large $J$). 
The most important difference between $J=0$ and $J\neq 0$ is 
a finite plateau value attained by correlation functions in the long time and large distances limit,
which is a residue of the coherence induced by the tunneling term.

Indeed, in the limit of large times and distances this correlation function tends to the asymptotic finite value:
\beq\label{CorrCxt2}
\displaystyle C(x \to \infty,t \to \infty) 
= \exp\Big[- \frac{1}{2 K}  {\mathcal I}_{\infty}  -  \frac{1}{8 K} \frac{\sqrt{\pi}}{2}   \Big] \ ,
\eeq
with:
\beq
\begin{array}{ll}
 \displaystyle  {\mathcal I}_{\infty} = \frac{\pi \, e^{\tilde{m}^2}}{2 \tilde{m}} \, {\rm Erfc}(\tilde{m})
 =  \frac{1}{\tilde{m}}  \Big[ \frac{\pi}{2}  + \mathcal{O}(\tilde{m})  \Big] 
 \ .
 \end{array}
\eeq

In the same semiclassical limit the thermal correlation function reads:
\beq\label{C_thermal}
 \displaystyle C_{\beta}(x) =     \exp\Big[ - \frac{1}{2 K } \displaystyle \int_0^{\infty} {\rm d} \tilde{p} \, 
 e^{- \tilde{p}^2} \frac{1}{\sqrt{\tilde{m}^2 + \tilde{p}^2}} (1 - \cos(\tilde{p} \tilde{x}) ) \, {\rm cotanh}\left(\frac{\tilde{\beta} \, \sqrt{\tilde{m}^2+\tilde{p}^2} }{2} \right)  \Big] \ ,
\eeq
where $\tilde{\beta}=\beta \hbar u/\xi_{h}$ and $\beta$ is the inverse temperature.
The thermal correlation function, at temperature $\beta_{\rm eff} =  \frac{4 K}{\rho \pi \hbar u}$ is shown in 
Fig. \ref{FIG2} with black solid lines.

\begin{figure}
  \begin{center}
        \includegraphics[scale=0.45]{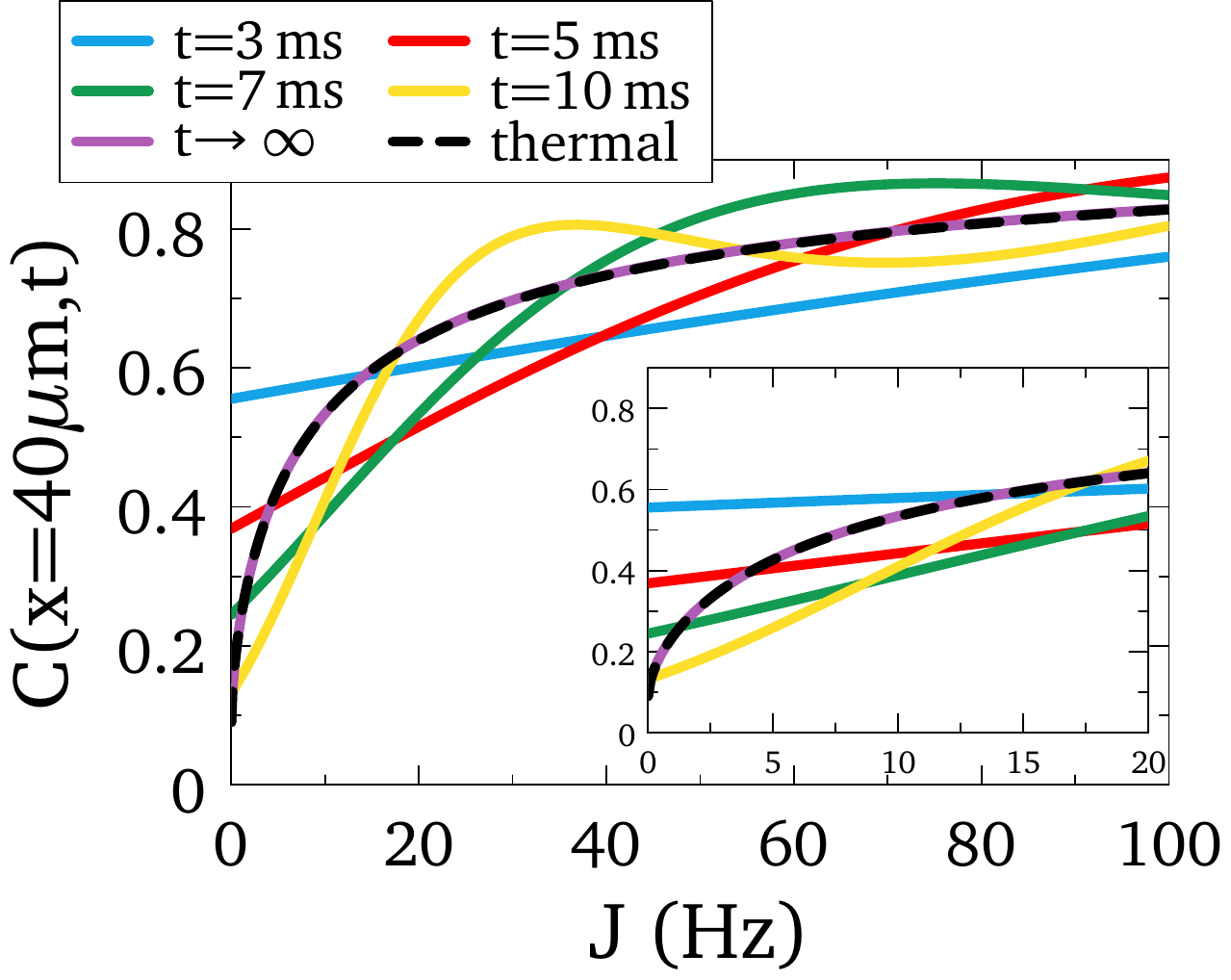}
                \includegraphics[scale=0.45]{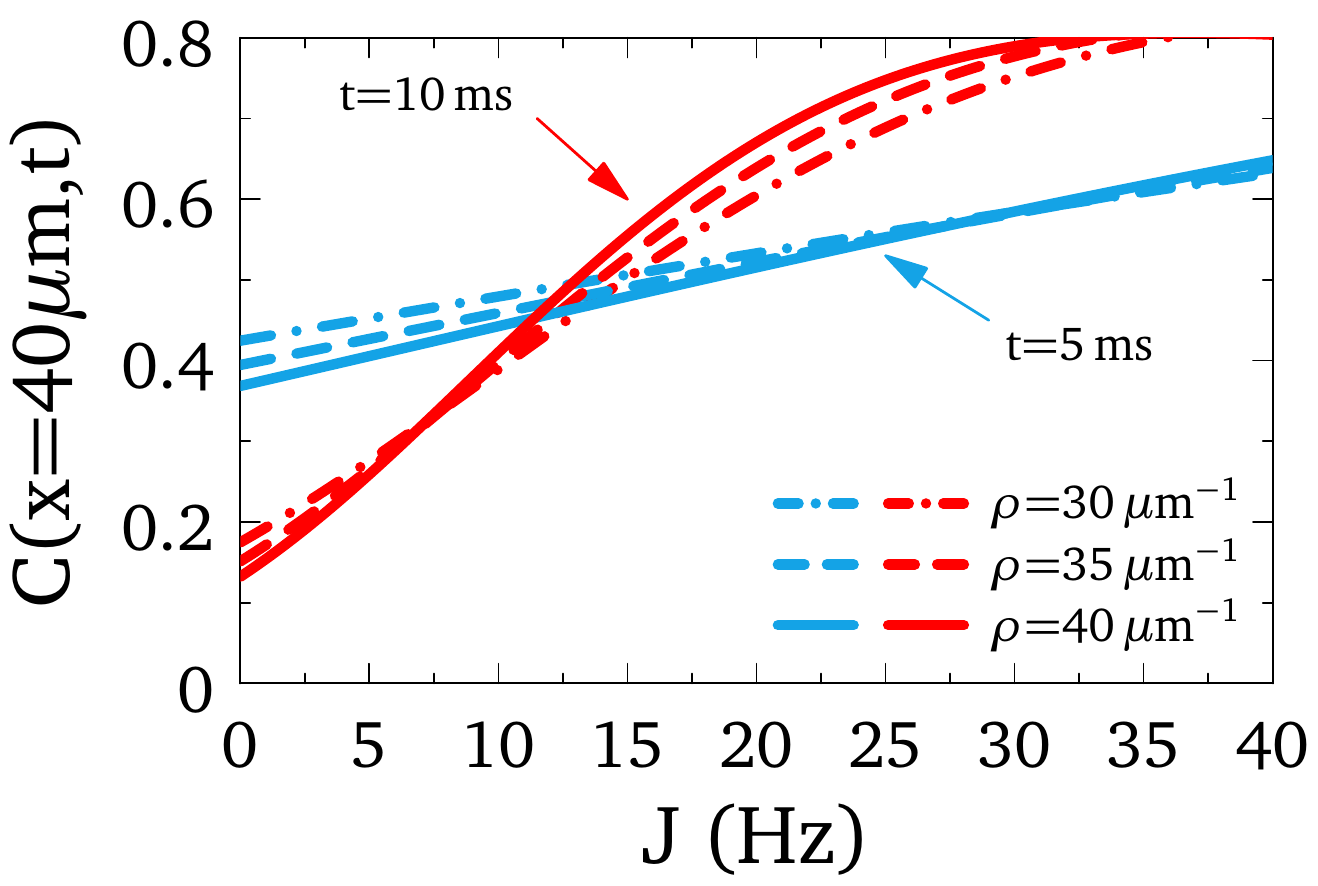}
        \caption{{\it Left panel:} $C(x=40\, \mu m,t)$ for $t=3, 5, 7,10$ ms  
        shown with blue, red, green and yellow curves respectively . The dashed black curve shows the value $C_{\beta}(x=40\, \mu m)$ given
       in Eq.~(\ref{C_thermal}). The violet curve (which is indistinguishable from the black one at thermal equilibrium)
       is the asymptotic stationary limit of (\ref{CorrCxt}). The inset highlights the region of small $J$ which is 
       more interesting for experiments.
       {\it Right panel:} Correlation functions at large distance as a function of the tunneling strength $J$ for different densities. 
        Blue and red lines refer respectively to times $t=5, 10$ ms. Dotted-dashed, dashed and solid
        lines refer to densities $\rho = 30, 35, 40 \ \mu m^{-1}$.
               }\label{FIG3}
  \end{center}
\end{figure}

In the left panel of Fig.~\ref{FIG3} we show the value of correlation functions at $x=40 \,\mu m$ as a function of $J$
for different times.
On the same Figure we show the thermal equilibrium result at $\beta_{\rm eff}$ which is indistinguishable 
from the stationary limit of Eq~(\ref{CorrCxt}).

Finally, the right panel of Fig.~\ref{FIG3} shows that the results are quite robust in response to variations of the density
of about $25\%$.

\section{Two-time correlation and response functions}\label{Sec_fdt}

An important way to test the thermalization of the system and eventually measure its 
temperature is to look at fluctuation-dissipation relations which relate equilibrium  
correlation and response functions via the temperature of the system~\cite{cugliandolo2011effective}.
From this perspective understanding how to access these two-time quantities 
experimentally in cold atomic experiments is a very interesting question~\cite{knap2013probing}.
Dynamical quantities at equilibrium for the Sine-Gordon model have been computed in~\cite{grits}.

In this section we consider the role of the effective temperature defined in (\ref{Teff}) 
at the level of the fluctuation-dissipation relation that involves the correlation and the response function of the relative phase.
In order to do this we compute the retarded and the Keldysh Green function of the relative phase~\cite{kamenev2011} 
where in the following 
we will denote $[A,B]=AB-BA$ and $\{A,B\}=AB+BA$. In particular we have:
\beq\label{GR}
\begin{array}{ll}
\displaystyle G_R(x,t ; x',t') 
& \displaystyle =  i \theta(t-t') \langle [\theta_A(x,t),\theta_A(x',t')] \rangle =
\\ \vspace{-0.2cm} \\
& \displaystyle =    \theta(t-t') 
\frac{1}{2\pi} \int {\rm d} p  \ e^{- i p (x-x') - \xi_h^2 p^2}\ \frac{\hbar \pi u}{\omega_p K}        \, \sin (\omega_{p}  (t -t'))   \ 
\end{array}
\eeq
Note that this response function does not contain information on the initial state, but only on the final Hamiltonian.
Moreover the Keldysh Green function reads:
\beq
\begin{array}{ll}
&\displaystyle G_K(x,t ; x',t') 
 \displaystyle = \frac12 \langle \{ \theta_A(x,t),\theta_A(x',t') \} \rangle = 
\\ \vspace{-0.2cm} \\
 & \displaystyle =  
 \frac{1}{4\pi} \int {\rm d} p \ e^{- i p (x-x') - \xi_h^2 p^2} 
  \ \Big[ ( \cos \omega_p (t-t') -  \cos \omega_p (t+t')  )  \ \left( \frac{\hbar\pi u}{\omega_p K}\right)^2  \langle |n_p|^2 \rangle(0) 
\\ \vspace{-0.2cm} \\
&  \displaystyle  \qquad \qquad +  ( \cos \omega_p (t-t') +  \cos \omega_p (t+t')  )   \  \langle |\theta_p|^2 \rangle(0)    \Big]
 \end{array}
\eeq
In the stationary limit the Keldysh Green function becomes:
\beq\label{GK_stat}
\begin{array}{ll}
\displaystyle G_K(x,t ; x',t') = \displaystyle  \frac{1}{4\pi} \int {\rm d} p  \ e^{- i p (x-x') - \xi_h^2 p^2}   \
 \Big[  \cos \omega_p (t-t')   \  \left( \frac{\hbar \pi u}{\omega_p K}\right)^2  \ \frac{\rho}{2}
\\ \vspace{-0.2cm} \\
\displaystyle
\qquad\qquad\qquad \qquad\qquad\qquad \qquad\qquad\qquad \qquad\qquad\
+    \cos \omega_p (t-t')  \  \frac{1}{2\rho}   \Big] \ .
 \end{array}
\eeq
The fluctuation-dissipation theorem (FDT) says that at equilibrium retarded and Keldysh Green function
are not independent and satisfy the following relation:
\beq
G_K (\omega) = {\rm cotanh} \frac{\omega}{2 T}  \ ( G_R (\omega) - G_A (\omega)) \ .
\eeq
In the classical limit the FDT simplifies and can be easily 
written in the time domain:
\beq
G_R (t) = - \frac{\theta(t)}{T} \partial_t G_K(t) .
\eeq
Therefore by comparing (\ref{GR}) and (\ref{GK_stat}) 
we note that
neglecting in (\ref{GK_stat}) the term $\rho^{-1}$, which corresponds to neglecting the initial fluctuation $\langle | \theta_p|^2\rangle(0)$ 
as compared to the relative density
one obtains:
 \beq\label{GK_stat2}
\begin{array}{ll}
\displaystyle \tilde{G}_K(x,t ; x',t') & \displaystyle = 
  \frac{1}{4\pi} \int {\rm d} p  \ e^{- i p (x-x') - \xi_h^2 p^2}     \cos \omega_p (t-t')   \ \left( \frac{\hbar \pi u}{\omega_p K}\right)^2  \ \frac{\rho}{2}   \ .
 \end{array}
\eeq
and the {\it classical} FDT is satisfied with the same $T_{\rm eff} = \frac{\hbar \pi u \rho}{4 K}$.

\section{Discussions}\label{Sec_discussions}

In section~\ref{Sec_dynamics_phase} we have shown that the presence of
a tunneling term, which in our case we treated via a semiclassical approximation
has several consequences on the dynamics of the relative phase
of the two gases with respect to the uncoupled case.

The transient regime and the way the steady state is reached are in fact 
 quite different than for the case where the two systems are independent. 
 One still sees a light-cone 
effect~\cite{LGKRS13,CC_2007,sotiriadis2010quantum}, manifested in Fig.~\ref{FIG2} 
as the length at which correlations show no further space dependences,
however after such characteristic (distance-dependent) time
the uncoupled gases acquire their stationary value while 
for $t_{\perp}\neq 0$ the correlations at short distances continue to evolve.
This can be interpreted as an effect due to the mass term 
which implies that not all the quasi-particles travel at the same velocity
and after the light-cone which is determined by the fastest excitations,
the correlations at short distances still present non stationary effects (as seen in Fig.~\ref{FIG2})
due to the slow modes.
This slowing down in reaching the steady state could make experimentally
difficult to observe stationary correlation functions.

As a consequence of the gapped spectrum one also observes in the
transient regime the occurrence of oscillations as a function of the space distance 
which can be interpreted as a non-equilibrium phenomenon 
due to the fact that the phase excitations exhibit a non-Lorentzian dispersion.

The other important difference with respect to the uncoupled case is that the correlation
functions of the relative phase display long-range order and do not decay. This is expected
because of the cosine term which favors minima in $\theta$ of multiples of $2\pi$.
In our approach we only considered the minimum around zero but
this effect should be robust if one takes into account solitons between different minima,
which are expected to weaken but not to destroy this order.
This result is also in agreement with~\cite{hofferberth2007non} where
long time phase coherence in presence of tunneling was also observed.

Despite these differences though, the effective temperature that emerges
from simple approximate arguments and by comparing with the asymptotic dynamics
does not depend on the tunneling strength and it is the same
that was found for the uncoupled case in~\cite{Gring_Science_2012,KISD11}.
The thermal correlation functions obtained with this temperature are
in fact indistinguishable from the ones obtained after the quench in the stationary limit
even if their precise analytical form differs.

These results suggest that in some cases a single temperature yields a quite accurate 
approximation of some
observables, even if the detailed dynamics is the one that
leads to a generalized Gibbs ensemble with an extensive number of
parameters.

However this is probably not a generic property of the post-quench dynamics
but a result of
how the splitting process creates the initial state. In fact by looking 
at the energy of the modes (\ref{Eq_Ek}) one sees that it is possible
to discard the dependences on the dispersion relation and thus both 
the dependencies on $k$ and on the tunneling strength
because one can neglect the initial fluctuations of the relative phase.
This is a key assumption that allows one to define a temperature 
equal for all the modes and one can see that the same approximation
allows us to recover the classical FDT in section \ref{Sec_fdt}.

In~\cite{LEGRSKRMGS2014} it was shown in fact that changing the quench protocol
drives the system in a state where a single temperature is not
sufficient to describe the steady state and instead more
Lagrange multipliers are needed to characterize accurately the
correlation functions of the interference measurement. 
It would be very
interesting to analyze the effect of the tunneling on this type of protocol as in that case one could expect that
the dependences on the dispersion relation and therefore on
the tunneling strength can modify more sensibly the Lagrange multipliers that
account for the constant of motion.
Moreover the possibility to measure two-time correlation and response functions
as we show in Section~\ref{Sec_fdt} for the prethermal state
would allow a direct measurement of these Lagrange multipliers~\cite{foini2016measuring}.

As a final remark let us note that in our analysis, as it was done in~\cite{Gring_Science_2012}
we assumed perfect decoupling between the symmetric and the antisymmetric mode.
This is the reason way the effective temperature that governs the dynamics
is not sensitive to the initial temperature of the system, which is instead
expected to appear to characterize the properties of the symmetric mode.
The account of the interaction between symmetric and antisymmetric mode
and the effect of the tunneling on this type of process and therefore on 
the final equilibration of the system
represents a very interesting perspective.

\section{Conclusions}\label{Sec_conclusions}

In this work we have studied the relaxation dynamics of two 
tunnel coupled gases created from the splitting on a single one dimensional
gas.

We have seen that the tunneling affects the non stationary dynamics of the
system with respect to the independent case leading to oscillations and a general
slowing down in reaching the steady state regime.

Differently from the uncoupled case the correlation functions in the steady state show 
long range coherence, however they are well described by the same effective temperature
that was measured in~\cite{Gring_Science_2012} in absence of tunneling.

The study of this type of protocol is nowadays possible with the current experimental
apparatus, and it would be therefore extremely interesting to test these consequences 
experimentally.

\section*{Acknowledgments}
We thank J. Schmiedmayer for interesting and very useful discussions.
This work was supported by the Swiss SNF under Division II, by the ARO-MURI Non-equilibrium Many-body 
Dynamics grant (W911NF-14-1-0003) and has received funding from the
European Research Council under the European Union's 7th Framework
Programme (FP/2007-2013/ERC Grant Agreement 307087-SPARCS).

\appendix

\section{Initial condition and cutoff}

In Sec.~\ref{Sec_initial_state} we assumed that the initial condition is 
of the form $\langle n_A(x) n_A(x')  \rangle(t=0) =  \frac{\rho}{2}  \, \delta(x-x')$.
The smearing of the delta function  over a scale of the order of the healing length $\xi_{h}$ 
can be written:
\beq
\langle n_A(x) n_A(x')  \rangle(t=0) = \frac{\rho}{4 \sqrt{\pi} \xi_h} e^{- \frac{(x-x')^2}{4\xi_h^2} } \ ,
\eeq
and similarly of $\theta_A$, whose amplitude fluctuations are encoded in
the parameter $(\xi_h \rho)^{-1} \propto K^{-1}$.
This initial condition provides a justification of the Gaussian cutoff that we used
in the computation of the correlation functions, in fact in Fourier space
it gives:
\beq
\langle n_A(p) n_A(p')  \rangle(t=0) = \delta_{p,-p'} \frac{\rho}{2} e^{-p^2 \xi_h^2} \ .
\eeq

\section{Computation of the correlation functions}\label{App2}

In order to prove Eq.~(\ref{CthetaA}) we follow \cite{KISD11}.
The initial state that gives rise to the initial correlation functions~(\ref{Initial_correlations})
can be written in the following squeezed form:
\beq\label{initialstate}
| \psi_0 \rangle = \frac{1}{N} \exp\big[ \sum_p W_p b^{\dag}_p b^{\dag}_{-p} \Big] |0\rangle |\psi_{p=0}\rangle \ ,
\eeq
where $ |0\rangle $ is the vacuum of bosons $b_p$, $W_p=\frac12 \frac{\pi\rho-|p|K}{\pi\rho+|p|K}$, $N$ is a normalization constant
and $ |\psi_{p=0}\rangle$ describes the component at $p=0$.
We note that one can define an operator $\gamma_p$ which annihilates the state (\ref{initialstate})
and this can be written as:
\beq
\gamma_p = - \frac{2 W_p}{\sqrt{1-W_p^2}} b_{-p}^{\dag} + \frac{1}{\sqrt{1-W_p^2}} b_{p} \ .
\eeq
Under the semiclassical approximation
the relative phase at any time can therefore be written as a linear combination 
of $\gamma_p$ and $\gamma_p^{\dag}$:
\beq
\theta_A(x,t)-\theta_A(0,t)  = \sum_{p} C_p(x,t) \gamma_p + C_{-p}^{\ast}(x,t) \gamma_{-p}^{\dag} \ .
\eeq
Using the identity $e^{A+B}=e^{A}e^{B}e^{-\frac12[A,B]}$, valid if $[A,B]$ is a $c$-number,
one gets:
\beq
\langle  \psi_0 | e^{i [\theta_A(x,t)-\theta_A(0,t) ]}   |  \psi_0 \rangle = e^{- \frac12  \sum_p | C_{p}(x,t) |^2 } \ .
\eeq
One also sees that:
\beq
\langle  \psi_0 | [\theta_A(x,t)-\theta_A(0,t)]^2 |  \psi_0 \rangle = \sum_p |C_{p}(x,t)|^2
\eeq
from which Eq.~(\ref{CthetaA}) follows.

\end{document}